\documentclass[twocolumn,tightenlines,showpacs,amsmath,amssymb,nofootinbib]{revtex4}
\usepackage{amssymb}
\usepackage{amsmath}
\usepackage{colordvi}
\usepackage{amsthm}
\usepackage{multirow}

\begin{document}
\newtheorem{theorem}{Theorem}
\newtheorem{lemma}{Lemma}
\newtheorem{corollary}{Corollary}

\title{Unambiguous discrimination of mixed states: A description based on system-ancilla coupling}
\author{Xiang-Fa Zhou}
\email{xfzhou@mail.ustc.edu.cn}
\author{Yong-Sheng Zhang}
\email{yshzhang@ustc.edu.cn}
\author{Guang-Can Guo}
\email{gcguo@ustc.edu.cn}
\affiliation{\textit{Key Laboratory of
Quantum Information, University of Science and Technology of
China, Hefei, Anhui 230026, People's Republic of China}}

\begin{abstract}
We propose a general description on the unambiguous discrimination
of mixed states according to the system-environment coupling, and
present a procedure to reduce this to a standard semidefinite
programming problem. In the two states case, we introduce the
canonical vectors and partly simplify the problem to the case of
discrimination between pairs of canonical vectors. By considering
the positivity of the two by two matrices, we obtain a series of
new upper bounds of the total success probability which depends on
both the prior probabilities and specific state structures.
\end{abstract}

\maketitle

\section{Introduction}
Quantum state discrimination (QSD) is one of the fundamentally
important problems in quantum information science. Especially in
quantum communication and quantum cryptography, many novel schemes
are based on the fact that nonorthogonal states cannot be
discriminated determinately. Henceforth study on the
discrimination of quantum states has a close relation to the
security of quantum cryptographic protocols. On the other hand,
since there is no measurement that can perform a perfect
identification, several strategies have been proposed in QSD based
on different criteria. One of these is the minimum-error
discrimination \cite{mim-error}, which permits incorrect outcomes
during the measurement procedure. The other one is unambiguous
discrimination (UD) of quantum states. This sort of discrimination
procedure never gives an erroneous result, but sometimes it fails.
Here we consider the latter case which has received much attention
recently.

In the pure states case, UD has been widely considered
\cite{early1,early2}. While in mixed states case, it seems to be a
hard problem. In many earlier works, some useful bounds of the
total success probability $P$, together with several useful
reduction theorem, have been presented \cite{mix-early, Rudolph,
mix-condition, Raynal1, Hergou, Raynal2,Raynal3, bergou}. However,
totally solving this problem seems not so easy at all. What's
more, even for the simple case, e.g., UD between two mixed states,
which has been wildly studied recently, there still exists many
questions which are not clear to us.

The standard description of UD among mixed states are usually
formulated as this: given a set of mixed states $\{\rho_1, \rho_2,
\ldots, \rho_N \}$ with the corresponding prior probabilities $\{
\eta_1, \eta_2, \ldots, \eta_N \}$, the aim of discriminating
these states unambiguously is to find a $N+1$-element positive
operator value measurement (POVM) $\{E_0, E_1, E_2, \ldots, E_N
\}$ with $\sum_{i=0}^N E_i=I$ and
$\mbox{Tr}(E_k\rho_l)=p_k\delta_{kl} (k,l\neq0)$, such that the
measurement operator $E_k$ gets a result with the probability
$p_k$ only when the input state is $\rho_k$. Here $E_0$ denotes
the inconclusive measurement where the identification fails. The
average failure probability is described by $Q=\sum_{k=1}^N Q_k$
with $Q_k=\eta_k \mbox{Tr}(E_0 \rho_k)$ being the failure
probability of identifying $\rho_k$. Equivalently, one can also
concentrate on the total success probability $P=1-Q=\sum_{k=1}
\eta_k \mbox{Tr}(E_k\rho_k)$.

From the general viewpoint, UD can be regarded as some kind of
physically accessible transformation on a finite number of input
states $\xi: \{\rho_1, \rho_2, \ldots, \rho_N
\}\rightarrow\{\sigma_1, \sigma_2, \ldots, \sigma_N \}$
\cite{uniform}. The major character of this transformation is that
it is probabilistic, accurate, and the output states $\sigma_k$
must be orthogonal to each other so that they can be identified
perfectly. There are several equivalent approaches to describe a
completely positive (CP) map \cite{book}. For example, it can be
represented in a Kraus operator sum form. Also it can be
implemented by employing a unitary transformation on the system
plus ancilla, i.e. $\xi(\rho)=\mbox{Tr}_{E'}[U \rho \otimes \rho_E
U^\dag I \otimes P_{E'}]$, where $\rho_E$ is the initial state  of
the ancilla system, $I$ denotes the identity operator in output
Hilbert space ${\cal H}_2$, $P_{E'}$ is a projector in ${\cal
H}_{E'}$, and ${\cal H}_1 \otimes {\cal H}_E = {\cal H}_2 \otimes
{\cal H}_{E'}$.

In this paper, we consider to discriminate mixed states
unambiguously from the system-ancilla coupling viewpoint. By
constructing the whole unitary transformation on the combinations
of the inputs and the auxiliary system, we obtain the necessary
and sufficient conditions on the existence of a UD strategy. We
point out that in the more general case to find the optimal UD
strategy can be reduced to a standard semidefinite programming
problem. Especially, in the case of UD between two mixed states,
we obtain a series of new upper bounds of the success probability
which are closely related to the structure of the input quantum
states together with the ratio of prior probabilities. In some
sense our result confirms the conjecture made by Bergou \emph{et
al} \cite{bergou}.

\section{General description of Unambiguous discrimination}
Let us start with the general UD of $N$ mixed states. For any
mixed state $\rho_k$, it can always be regarded as the mixture of
pure states, i.e. $\rho_k=\sum_m|\tilde{\psi}_m^{(k)}\rangle
\langle \tilde{\psi}_m^{(k)}|$, where
$|\tilde{\psi}_m^{(k)}\rangle$ are nonnormalized state vectors.
Here for simplicity, we assume that $|\tilde{\psi}_m^{(k)}\rangle$
are linearly independent. Firstly, we suppose the intersection of
the supports of two density matrices $\rho_k$ and $\rho_l$ is
empty (except for a trivial zero vector). This also indicates that
all vectors $\{|\tilde{\psi}_m^{(k)}\rangle, \ldots,
|\tilde{\psi}_n^{(l)}\rangle \}$ are linearly independent. By
introducing suitable auxiliary system, we consider the following
unitary realization of the CP map $\xi$
\begin{eqnarray}
U |\tilde{\psi}^{(k)}_{m} \rangle_1 |0\rangle_E =
|\tilde{\phi}^{(k)}_{m} \rangle_{2a} |P_0\rangle_p +
|\tilde{\beta}^{(k)}_{m} \rangle_{2ap}.
\end{eqnarray}
Here ${\cal H}_{E'}={\cal H}_{a} \otimes {\cal H}_{p}$,
$|0\rangle$ is the fixed initial state of the environment,
$|P_0\rangle$ is the state of the probe system satisfying $\langle
P_0|\widetilde{\beta}^{(k)}_{m}\rangle=0$, and we also use the
tilde $\widetilde{\mbox{ }}$ to denote a nonnormalized state
vector. The output state $\sigma_k$ can be obtained by tracing
over the subsystem $a$ after we get a measurement outcome
corresponding to the probe $|P_0\rangle$, i.e.
$\sigma_k=\sum_m\mbox{Tr}_a[|\tilde{\phi}^{(k)}_{m} \rangle
\langle \tilde{\phi}^{(k)}_{m}| ]$.

On the other hand, if the intersection of the supports of two
density matrix $\rho_k$ and $\rho_l$ is not empty, there exists at
least one state vector $|\psi\rangle \in supp(\rho_k) \cap
supp(\rho_l)$. From the definition of the CP map, we have
\begin{eqnarray}
U |\psi\rangle |0\rangle &=& |\tilde{\phi}^{(k)}\rangle
|P_0\rangle + |\tilde{\beta}^{(k)} \rangle \nonumber \\
&=& |\tilde{\phi}^{(l)}\rangle |P_0\rangle +
|\tilde{\beta}^{(l)}\rangle. \label{supprot}
\end{eqnarray}
Since $|\tilde{\phi}^{(k)}\rangle \neq |\tilde{\phi}^{(l)}
\rangle$ (The output states are different from each other), Eq.
($\ref{supprot}$) is satisfied only when
$|\tilde{\phi}^{(k)}\rangle = |\tilde{\phi}^{(l)} \rangle=0$,
hence any state contained in $supp(\rho_k) \cap supp(\rho_l)$ has
no contribution to the desired transformation. Thus it's enough to
consider the case $supp(\rho_k) \cap supp(\rho_l)= \{ 0 \}$, which
reproduces the known results \cite{mix-condition}.

The inner-product preservation of unitary transformation leads us
to the following equation
\begin{eqnarray}
\tilde{X}-\tilde{Y}=\tilde{B} \ge 0 \label{main}
\end{eqnarray}
with
\begin{eqnarray}
\tilde{w}=
\left ( \begin{array}{ccc} \tilde{w}_{kk} & \ldots &  \tilde{w}_{kl} \\
\vdots & \ddots & \vdots
\\  \tilde{w}_{lk} & \cdots & \tilde{w}_{ll}
\end{array} \right )  \,\,\,\, \{w \in (X, Y, B) \}.
\end{eqnarray}
Here $\tilde{w}_{kl}$ are all block matrices with
$(\tilde{X}_{kl})_{mn}=\langle
\tilde{\psi}^{(k)}_m|\tilde{\psi}^{(l)}_n \rangle$,
$(\tilde{Y}_{kl})_{mn}=\langle
\tilde{\phi}^{(k)}_m|\tilde{\phi}^{(l)}_n \rangle$, and
$(\tilde{B}_{kl})_{mn}=\langle
\tilde{\beta}^{(k)}_m|\tilde{\beta}^{(l)}_n \rangle$ respectively.
Also we can find that all the three matrices ($\tilde{X},
\tilde{Y}, \tilde{B}$) are Hermitian, and positive semidefinite.
Since $\sigma_k$ are orthogonal to each other, we have $\langle
\tilde{\phi}^{(k)}_{m}| \tilde{\phi}^{(l)}_{n} \rangle=0 (k\neq
l)$. This indicates $\tilde{Y}$ is quasi-diagonal and can be
written as $ \tilde{Y}= diag\{\tilde{Y}_{kk}, \ldots,
\tilde{Y}_{ll} \}$.

Contrarily, if there exists a positive semidefinite $\tilde{Y}$
matrix satisfying Eq. ($\ref{main}$), we can always choose
suitable state vectors $|\tilde{\phi}^{(k)}_{m} \rangle$ and
$|\tilde{\beta}^{(k)}_{m} \rangle$ such that
$\tilde{X}=\tilde{B}+\tilde{Y}$. With the standard Gram-Schmidt
orthogonalization procedure, the desired the unitary
transformation can be easily obtained.

We conclude the above discussion by the following theorem.
\begin{theorem}
$N$ mixed states  $\{\rho_1, \rho_2, \cdots \rho_N \}$ can be
unambiguously discriminated if and only if there exists a positive
semidefinite quasi-diagonal matrix $\tilde{Y}$ such that
$\tilde{X}- \tilde{Y} \geq 0$. Moreover, if the input states are
chosen with prior probabilities $\{ \eta_1, \eta_2,\cdots
\eta_N\}$ and $\sum_k \eta_k=1$, the total success probability
will be $P=\sum_k \eta_k \mbox{Tr}(\tilde{Y}_{kk})$.
\end{theorem}

This theorem characterize the general properties of UD among $N$
mixed states in the system-ancilla framework. One can also easily
check that it is consistent with earlier works \cite{mix-early,
Rudolph, mix-condition, Raynal1, Hergou, Raynal2,Raynal3, bergou}.
In a more realistic situation, people often concentrate on the
total success probability of such physical transformation. This
indicates that we should make the probability $P$ as high as
possible. Mathematically, this is equivalent to maximizing
\begin{eqnarray}
P=\sum_k \eta_k \mbox{Tr}(\tilde{Y}_{kk})
\end{eqnarray}
under the constraints
\begin{eqnarray} \label{sdp2}
\tilde{X} - \tilde{Y} \geq 0, and \ \tilde{Y} \geq 0.
\end{eqnarray}
Usually given the input mixed states, we can get to know the
matrix $\tilde{X}$ exactly. Therefore the only thing we should do
is to find the optimal positive semidefinite matrix $\tilde{Y}$
which maximizes the success probability $P$. By redefining a
series of new matrices $ F_0=diag\{\tilde{X},0 \}$, $ F_{k}^{pq} =
diag\{ E_k^{pq}, -E_k^{pq}\}$, and $G_k^{pq} = diag \{i E_k^{pq},
-i E_k^{pq} \}$, where $i$ is the basic imaginary unit, and
$E_k^{pq}$ are matrices corresponding the block matrices
$\tilde{Y}_{kk}$ with $(E_k^{pq})_{mn}=\delta_{mp}\delta_{nq}$,
the problem under consideration can be reformulated as
\begin{eqnarray}
{\hbox{\space \raise-2mm\hbox{$\textstyle max \atop \scriptstyle
\tilde{Y}$} \space}} \sum_k \eta_k \mbox{Tr}(\tilde{Y}_{kk})
\end{eqnarray}
subject to
\begin{eqnarray}\label{sdp} \nonumber
F_0 &-& \sum_{k,p,q } \left \{ Re[(\tilde{Y}_{kk})_{pq}] F_k^{pq}
+ Im[(\tilde{Y}_{kk})_{pq}] G_k^{pq} \right \}  \geq 0,
\end{eqnarray}
where $Re[(\tilde{Y}_{kk})_{pq}]$ and $Im[(\tilde{Y}_{kk})_{pq}]$
represent the real and imaginary parts of the matrices elements
$(\tilde{Y}_{kk})_{pq}$ respectively. This is a standard
semi-definite programming (SDP) problem \cite{SDP}, and can be
solved by numeric method efficiently (one can also find another
method to reduce this to a SDP problem in \cite{sdp2}, which is
equivalent to our result). Therefore in principle, the optimal
success probability of UD of mixed states can be found
numerically. Actually once we have found the optimal matrix
$\tilde{Y}$, with the standard procedure, we can construct the
corresponding unitary implementation of the discrimination
operation.

\section{Unambiguous discrimination of two mixed states}
In the above discussion, we have given a general description on UD
among $N$ mixed input states. To be specific, in the following, we
will focus on a particular case, i.e. UD between two mixed states.
This is a basic and very important case in the study of UD, and
much attention has been paid to the problem recently. In
\cite{Rudolph}, Rudolph \emph{et al.} present the lower bound on
the failure probability $Q$, and later, it has been pointed out
that there exist mixed states for which the lower bound can not be
reached for any prior probabilities. Based on these facts, Raynal
\emph{et al.} \cite{Raynal2, Raynal3} investigated a large class
of two mixed states discrimination, they also found the necessary
and sufficient conditions for two mixed state to saturate these
bounds. In all these works, $Q$ is considered in three different
regions, depending on the ratio between two prior probabilities.
Recently, Bergou \emph{et al.} \cite{bergou} have considered the
discrimination of two subspaces and they find that for this
special case there are many parameter regions which can give
different minimal failure probabilities. The regions depend on
both the prior probabilities and the specific structure of the two
subspaces. The lower bound $2\sqrt{\eta_1 \eta_2}F$ of the failure
probability $Q$ can be reached only when the prior probabilities
lie in some specific regions. Later they conjecture that this
phenomenon occurs for any two mixed states. In the following, we
will show that this result is indeed universal.

When restricted to two-state case, Eq. ($\ref{main}$) can be
simplified as
\begin{eqnarray} \label{two}
\left ( \begin{array}{cc} \tilde{X}_{11} & \tilde{X}_{12} \\
  \tilde{X}_{21} & \tilde{X}_{22} \end{array} \right ) -
\left ( \begin{array}{cc} \tilde{Y}_{11} & 0 \\
  0 & \tilde{Y}_{22} \end{array} \right ) \ge 0,
\end{eqnarray}
where $\tilde{X}_{kl}$ arise from the decompositions of $\rho_1$
and $\rho_2$. Usually there exist many other ensembles which can
generate the same operators, i.e.
$\rho_1=(|\tilde{\psi}_1^{(1)}\rangle,
|\tilde{\psi}_2^{(1)}\rangle, \ldots)(\langle
\tilde{\psi}_1^{(1)}|, \langle \tilde{\psi}_2^{(1)}|,
\ldots)^T=(|\tilde{\psi}_1^{(1)}\rangle,
|\tilde{\psi}_2^{(1)}\rangle, \ldots) U^\dag U  (\langle
\tilde{\psi}_1^{(1)}|, \langle \tilde{\psi}_2^{(1)}|,
\ldots)^T=(|\tilde{r}_1\rangle, |\tilde{r}_2\rangle,
\ldots)(\langle \tilde{r}_1|, \langle \tilde{r}_2|, \ldots)^T$,
where $U$ ($V$ for $\rho_2$) is a unitary matrix and $T$
represents the transpose of the matrix. This is known as the
unitary freedom for density matrices \cite{book2}. Hence we can
also write down the correspondence of Eq. ($\ref{main}$) according
to this new decomposition
\begin{eqnarray} \label{two-2}
\tilde{X'}-\tilde{Y'} \ge 0.
\end{eqnarray}
Since
$\tilde{X'}=\mbox{diag}\{U,V\}\tilde{X}\mbox{diag}\{U^\dag,V^\dag\}$,
we can immediately obtain that this will not affect the total
success probability $P$ we consider here (This is also general for
$N$ input mixed states).

Keeping in mind that $U$ and $V$ can be arbitrary, we can choose
the two matrices appropriately such that $U \tilde{X}_{12} V^\dag=
\mbox{diag}\{\mbox{diag}\{ f_1, f_2, \ldots, f_t \},
\overrightarrow{\textbf{0}} \}$, where we assume $\tilde{X}_{11}$
and $\tilde{X}_{22}$ are $u\times u$ and $v\times v$ matrices
respectively with $u\le v$, $f_m$ are the singular values of
$X_{12}$, and $\overrightarrow{\textbf{0}}$ is a $(u-t)\times
(v-t)$ zero matrix. This implies there exist some kinds of
decompositions of $\rho_1$ and $\rho_2$, namely,
$\rho_1=\sum_m|\tilde{r}_m\rangle \langle \tilde{r}_m|$ and
$\rho_2=\sum_n|\tilde{s}_n\rangle \langle \tilde{s}_n|$, which
satisfies the following equations
\begin{eqnarray} \label{vector}
\langle \tilde{r}_m| \tilde{s}_n \rangle = \left \{
\begin{array}{ll}
f_m \delta_{mn}  &   (m,n) \le t, \\
0                &   otherwise. \end{array} \right.
\end{eqnarray}
The singular values $f_m$ have very interesting properties and we
characterize this by the following theorem \cite{Uhlmann}.
\begin{theorem}
Given two mixed states density matrices $\rho_1$ and $\rho_2$,
there exist two sets of canonical vectors $\{|\tilde{r}_1\rangle,
|\tilde{r}_2\rangle, \ldots \}$ and $\{|\tilde{s}_1\rangle,
|\tilde{s}_2\rangle, \ldots \}$, which generate $\rho_1$ and
$\rho_2$ respectively, such that Eq. ($\ref{vector}$) is
satisfied. And the fidelity of the two density matrices can be
formulated as
$F=\mbox{Tr}\sqrt{\rho_1^{1/2}\rho_2\rho_1^{1/2}}=\sum_m f_m$.
\end{theorem}
\emph{Proof}: The only thing we should do now is to prove the
second part of this theorem. Consider the spectral decompositions
of $\rho_1$ and $\rho_2$
\begin{eqnarray}
\rho_1=\sum_i \alpha_i |\alpha_i \rangle \langle \alpha_i|,
\,\,\,\, \rho_2=\sum_h \beta_h |\beta_h \rangle \langle \beta_h|.
\end{eqnarray}
According to the definition of fidelity $F$, we obtain
\begin{eqnarray}
(\rho_1^{1/2}\rho_2\rho_1^{1/2})_{ij}=& \sum_h &
\sqrt{\alpha_i}\sqrt{\beta_h}\langle \alpha_i|\beta_h\rangle
 \nonumber \\ &\mbox{}&
\!\!\cdot \sqrt{\alpha_j}\sqrt{\beta_h}\langle \beta_h|\alpha_j
\rangle.
\end{eqnarray}
Now we definite a new matrix
\begin{eqnarray}
A_{ih} = \sqrt{\alpha_i}\sqrt{\beta_h}\langle
\alpha_i|\beta_h\rangle.
\end{eqnarray}
The fidelity $F$ can be rewritten as this
\begin{eqnarray}
F=\mbox{Tr}\sqrt{\rho_1^{1/2}\rho_2\rho_1^{1/2}}=\mbox{Tr}\sqrt{AA^\dag}.
\end{eqnarray}
Since $A$ is a complex matrix, using a singular value
decomposition, we have $A=U_1 \mbox{diag}\{\mbox{diag}\{ f_1, f_2,
\ldots, f_t \}, \overrightarrow{\textbf{0}} \} V_1$ with $f_m \ge
0$ for all $1\le m \le t$. Thus the fidelity becomes
\begin{eqnarray}
F=\mbox{Tr}\sqrt{AA^\dag}=\sum_m f_m.
\end{eqnarray}
On the other hand, because of the unitary freedom in the ensemble
representation of density matrices, we can find two unitary
operations $U_2$ and $V_2$ such that $U_2 X_{12} V_2=A$. Therefore
$A$ and $X_{12}$ have the same singular values, which completes
the proof.

Theorem 2 indicates for any two mixed states, it is always
possible to find two sets of canonical vectors
$\{|\tilde{r}_1\rangle, |\tilde{r}_2\rangle, \ldots \}$ and
$\{|\tilde{s}_1\rangle, |\tilde{s}_2\rangle, \ldots \}$ so that
$|\tilde{r}_m\rangle$ only have a nonzero overlap with
$|\tilde{s}_m\rangle$. When $(n, m) \ge t$, one can easily check
that $|\tilde{r}_m\rangle$ and $|\tilde{s}_n\rangle$ lie in the
subspace orthogonal to the supports of $\rho_2$ and $\rho_1$
respectively. From the reduction theorem in \cite{Raynal1}, we
conclude that UD between $\rho_1$ and $\rho_2$ is equivalent to
that between the two newly defined density matrices
$\rho_1'=\sum_{m=1}^t |\tilde{r}_m\rangle \langle
\tilde{r}_m|/N_1$ and $\rho_2'=\sum_{n=1}^t |\tilde{s}_n\rangle
\langle \tilde{s}_n|/N_2$ with $N_1=\mbox{Tr}(\sum_{m=1}^t
|\tilde{r}_m\rangle \langle \tilde{r}_m|)$ and
$N_2=\mbox{Tr}(\sum_{n=1}^t |\tilde{s}_n\rangle \langle
\tilde{s}_n|)$ being the corresponding normalization factors.
According to the system-ancilla model (Theorem 1), Eq.
($\ref{two}$) can always be reduced to a $2t \times 2t$ matrix
\begin{eqnarray}\label{three}
\left ( \begin{array}{cc} \tilde{X}_{11}-\tilde{Y}_{11} & \mbox{diag}\{f_1, \ldots, f_t\} \\
  \mbox{diag}\{f_1, \ldots, f_t\} & \tilde{X}_{22}-\tilde{Y}_{22} \end{array} \right )  \ge
  0,
\end{eqnarray}
where we have used the same notations for simplicity.

Equation ($\ref{three}$) supplies enough information which can be
used to demonstrate our main results. Actually, since
$\tilde{X}-\tilde{Y}$ is positive semidefinite, from the standard
linear algebra theory, we have that every principal minor of
$\tilde{X}-\tilde{Y}$ is also positive semidefinite, i.e.
\begin{eqnarray}
\left ( \begin{array}{cc} r_m-y_m &  f_m \\
  f_m & s_m-z_m \end{array} \right )  \ge
  0,
\end{eqnarray}
where we have used $y_m$ and $z_m$ to denote the diagonal elements
of $\tilde{Y}_{11}$ and $\tilde{Y}_{22}$ respectively, and $r_m$
($s_m$) is the modulus of the vector $|\tilde{r}_m \rangle$
($|\tilde{s}_m \rangle$). Therefore by introducing the canonical
state vectors, the question can be in part reduced to UD between
pairs of state vectors $|\tilde{r}_m \rangle$ and
$|\tilde{s}_m\rangle$. Such question has been solved in many
earlier works, and the results are listed as follows
\begin{eqnarray} \label{newbound}
P_m \!\!\! &=& \!\! \eta_1 y_m + \eta_2 z_m \nonumber \\
    &\le& P_m^{max}\nonumber \\
    & = &\!\! \left \{
    \begin{array}{ll}
    \eta_2(s_m - f_m^2/r_m)  &  0 \le \sqrt{\frac{\eta_1}{\eta_2}} \le
    \frac{f_m}{r_m}, \\
    \eta_1 r_m + \eta_2 s_m -2\sqrt{\eta_1 \eta_2}f_m  & \frac{f_m}{r_m} \le
    \sqrt{\frac{\eta_1}{\eta_2}} \le \frac{s_m}{f_m}, \\
    \eta_1(r_m - f_m^2/s_m)  &  \sqrt{\frac{\eta_1}{\eta_2}} \ge
    \frac{s_m}{f_m}.
    \end{array}
    \right .\nonumber\\
    \mbox{                        }
\end{eqnarray}

The above expression shows that for every $m$, the maximal value
that $P_m$ can achieve has a close relation with the specific
configuration of $\sqrt{\frac{\eta_1}{\eta_2}}$, $r_m$, $s_m$, and
$f_m$. Generally, for different $m$, $P_m$ will have very
different expressions. Therefore, the total success probability
$P=\sum_mP_m$ cannot always be represented as a function of the
fidelity $F=\sum_m f_m$.

Specifically, in the following we will focus on some special
cases. Firstly, if for all $m=1, \ldots, t$, we have
$\frac{f_m}{r_m} \le
    \sqrt{\frac{\eta_1}{\eta_2}} \le \frac{s_m}{f_m}$, then
according to the above equation, the upper bound of the total
success probability can be rewritten as
\begin{eqnarray}
P &=& \sum_m P_m \le \sum_m P^{max}_m \nonumber \\
    &=& \sum_m \eta_1 r_m + \eta_2 s_m - 2 \sqrt{\eta_1 \eta_2}
    f_m \nonumber \\
    &=& 1 - 2 \sqrt{\eta_1 \eta_2} F.
\end{eqnarray}
The corresponding lower bound of the failure probability becomes
$Q=1-P \ge 2 \sqrt{\eta_1 \eta_2}F$. This bound has been proved to
be the minimal value of $Q$ for any type of input configurations.
However, our result shows that even in this special case, the
lower bound of $Q$ can only be possibly saturated. This occurs,
for example, when the canonical vectors are orthogonal to each
other. In general case, since $\tilde{X}_{11}$ and
$\tilde{X}_{22}$ are not diagonal matrices, this lower bound
cannot always be reached.

In the second example, we assume that
$\sqrt{\frac{\eta_1}{\eta_2}} \ge \frac{s_m}{f_m}$ for all $1 \le
m \le t$. A simple algebra will lead us to the following bound of
the total success probability $P  \le \eta_1(1-\sum_m f_m^2/s_m)$.
 If we introduce a new operator $C_2=\sum_m
|s_m\rangle \langle s_m|$ composed of the normalized canonical
vectors of $\rho_2$, we can reformulate $P $ as $P  \le
\eta_1(1-\mbox{Tr}(\rho_1 C_2))$, or equivalently $Q  \ge \eta_2 +
\eta_1 \mbox{Tr}(\rho_1 C_2)$. When $|s_m\rangle$ are orthogonal
to each other, $C_2$ is nothing but the projection onto the
support of $\rho_2$.

Thirdly, if we have $\sqrt{\frac{\eta_1}{\eta_2}} \le
\frac{f_m}{r_m} (\forall m)$, the total success probability
satisfies $P \le \eta_2(1-\sum_m
f_m^2/r_m)=\eta_2(1-\mbox{Tr}(\rho_2 C_1))$ with $C_1=\sum_m
|r_m\rangle \langle r_m|$. Correspondingly, the failure
probability becomes $Q \ge \eta_1 + \eta_2 \mbox{Tr}(\rho_2 C_1)$.

For mixed states $\rho_1$ and $\rho_2$, we always have $(r_m,
s_m)<1$. This indicates that the failure probability $Q$ can never
reach the bound $2\sqrt{\eta_1 \eta_2}F$ for the latter two cases.
Generally, different canonical vectors of the input states will
separate the parameter space into different regions, and the lower
bound of $Q$ is determined by both the prior probabilities and the
structure of states. Moreover, in each region, the lower bound of
$Q$ can not always be reached. Mathematically, to judge whether
the lower bound can be saturated is equivalent to determining
whether there exists a positive semidefinite matrix $\tilde{Y} \ge
0$ such that Eq. ($\ref{sdp}$) is satisfied. This problem is often
called semidefinite feasibility problem (SDFP). Unfortunately, the
complexity of SDFP is still not known, and currently we can only
say that it cannot be a NP-complete problem unless NP=NP-complete.
Therefore to judge whether the bound of $Q$ can be reached or not
seems to be a hard problem. But in some special case (for example,
the canonical vectors are orthogonal to each other, or
$\tilde{X}-\tilde{Y}$ is a diagonally dominant matrix), some known
results in linear algebra theory will be helpful to solve this
problem.

\section{examples}
In many related works, the upper bound of the success probability
$P$ is only considered in three different intervals, which depends
on the ratio of $\eta_1$ and $\eta_2$ together with the fidelity
$F$ and supports of the input states. Here by introducing the
decomposition in Theorem 2, we find a series of parameter regions
related to the specific input states. In addition, from the
system-ancilla coupling viewpoint, one can also derive the
corresponding results in \cite{Raynal2} and \cite{Rudolph}. For
example, if we definite $\tilde{B} = \tilde{X'}-\tilde{Y'}$ in Eq.
($\ref{two-2}$), then since $\tilde{B}$ is positive semidefinite,
we have $\sqrt{\mbox{Tr}(\tilde{B}_{11})\mbox{Tr}(\tilde{B}_{22})}
\ge |\mbox{Tr}(\tilde{B}_{12})|$ for any kind of decompositions of
$\rho_1$ and $\rho_2$ (for the definitions of $\tilde{B_{ij}}$,
see Eq. ($\ref{main}$)). Therefore we obtain
$\sqrt{\mbox{Tr}(\tilde{B}_{11})\mbox{Tr}(\tilde{B}_{22})} \ge F$,
where equality holds only when $\tilde{B}_{11}=\alpha
\tilde{B}_{22}$ with $\alpha \in \mathbb{R}$. This also indicates
that the output states corresponding to the failure measurement
results cannot be used for further discrimination operations,
which is consistent with the discussions in \cite{Raynal2}. To
reveal the relation and difference between the bounds list above
and those in the previous works, in the following we will
investigate a specific example.

Consider two rank-$2$ mixed states
$\rho_1=\frac{1}{2}(|r_1\rangle\langle r_1|+|r_2\rangle \langle
r_2|)$ and $\rho_2=\frac{1}{2}(|s_1\rangle \langle
s_1|+|s_2\rangle \langle s_2|)$ with $\langle
r_1|s_2\rangle=\langle r_2|s_1\rangle=0$, $\langle
r_1|s_1\rangle=\mbox{cos}\theta_1$, and $\langle
r_2|s_2\rangle=\mbox{cos}\theta_2$. To simplify our consideration,
we also assume $\langle r_1|r_2\rangle=\langle s_1|s_2\rangle=0$.
Actually, discrimination of such kind of mixed states has be
extensively studied in \cite{bergou}. Here we also use it to
manifest the difference of the upper bounds presented in several
related works. Suppose $0 < \mbox{cos}\theta_1 \le
\mbox{cos}\theta_2 <1$. Then based on our former discussions,
optimal success probability $P$ can be obtained exactly in five
parameter regions, $[0, \mbox{cos}\theta_1]$,
$[\mbox{cos}\theta_1, \mbox{cos}\theta_2]$, $[\mbox{cos}\theta_2,
1/\mbox{cos}\theta_2]$, $[1/\mbox{cos}\theta_2,
1/\mbox{cos}\theta_1]$, $[1/\mbox{cos}\theta_1, \infty)$.
Alternatively, one can also obtain the corresponding upper bound
of $P$ according to \cite{Raynal2} and \cite{Rudolph}. Table 1
shows the details of $P$ in each of these regions.

\begin{widetext}

\begin{table}\label{table-I}
\begin{center}
\begin{tabular}{||c||c|c|}\hline
$\sqrt{\eta_1/\eta_2}$ & $P$ & $P_{Ra}(P_{Ru})$ \\ \hline $[0,
\mbox{cos} \theta_1]$ & $\frac{1}{2}\eta_2(\mbox{sin}^2 \theta_1 +
\mbox{sin}^2 \theta_2)$ & $\frac{1}{2}\eta_2(\mbox{sin}^2\theta_1
+ \mbox{sin}^2 \theta_2)+ $
\\ \cline{1-2}
$[\mbox{cos}\theta_1, Ra_1(F)]$ & \multirow{2}{*}{$\frac{1}{2}-\sqrt{\eta_1 \eta_2}\mbox{cos}\theta_1+\frac{1}{2}\eta_2 \mbox{sin}^2\theta_2$} & $\frac{1}{2}\eta_1\frac{(\mbox{cos}\theta_1-\mbox{cos}\theta_2)^2}{\mbox{cos}^2\theta_1+\mbox{cos}^2\theta_2}$ (or $\eta_2(1-F^2)$)  \\
\cline{1-1} \cline{3-3} $[Ra_1(F), \mbox{cos}\theta_2]$ & &
\multirow{3}{*}{$1-2 \sqrt{\eta_1 \eta_2}F$  (or $1-2\sqrt{\eta_1
\eta_2}F$) } \\ \cline {1-2} $[\mbox{cos}\theta_2,
\frac{1}{\mbox{cos}\theta_2}]$ & $1-2\sqrt{\eta_1 \eta_2}F$&  \\
\cline{1-2} $[\frac{1}{\mbox{cos}\theta_2}, Ra_2(1/F)]$
&\multirow{2}{*}{$\frac{1}{2}-\sqrt{\eta_1
\eta_2}\mbox{cos}\theta_1+ \frac{1}{2}\eta_1 \mbox{sin}^2\theta_2$} & \\
\cline{1-1} \cline{3-3}
$[Ra_2(1/F), \frac{1}{\mbox{cos}\theta_1}]$ &  & $\frac{1}{2}\eta_1(\mbox{sin}^2\theta_1+\mbox{sin}^2\theta_2)+$ \\
\cline{1-2} $[\frac{1}{\mbox{cos}\theta_1}, \infty)$ &
$\frac{1}{2} \eta_1 (\mbox{sin}^2\theta_1+\mbox{sin}^2\theta_2)$ &
$\frac{1}{2}\eta_2
\frac{(\mbox{cos}\theta_1-\mbox{cos}\theta_2)^2}{\mbox{cos}^2\theta_1+\mbox{cos}^2\theta_2}$
(or $\eta_1(1-F^2)$) \\ \hline
\end{tabular}
\end{center}
\caption{Bounds of the maximal success probabilities presented in
several related works. Here $P$ denotes the bound according to Eq.
($\ref{newbound}$). $P_{Ra}$ and $P_{Ru}$ are the results obtained
from \cite{Raynal2} and \cite{Rudolph} respectively.
$F=(\mbox{cos}\theta_1+\mbox{cos}\theta_2)/2$ is the fidelity of
the two input mixed states. $Ra_1=\mbox{Tr}(P_1
\rho_2)/F=(\mbox{cos}^2\theta_1+\mbox{cos}^2\theta_2)/(\mbox{cos}\theta_1+\mbox{cos}\theta_2)$
and $Ra_2=F/\mbox{Tr}(P_2
\rho_1)=(\mbox{cos}\theta_1+\mbox{cos}\theta_2)/(\mbox{cos}^2\theta_1+\mbox{cos}^2\theta_2)$
are parameters according to \cite{Raynal2} with $P_1$ and $P_2$
being the supports of $\rho_1$ and $\rho_2$ separately.}
\end{table}
\end{widetext}

The above table shows that when
$\mbox{cos}\theta_1=\mbox{cos}\theta_2$, the three bounds $P$,
$P_{Ra}$, and $P_{Ru}$ are equal to each other. However, for the
general case $\mbox{cos}\theta_1 \ne \mbox{cos}\theta_2$, one can
easily obtain $P \le P_{Ra}$ and $P \le P_{Ru}$, and equalities
hold only when $\mbox{cos}\theta_2 \le \sqrt{\eta_1/\eta_2} \le
1/\mbox{cos}\theta_2$. For example, if $\mbox{cos}\theta_1 \le
\sqrt{\eta_1/\eta_2} \le Ra_1$, we have
$P_{Ra}-P=\frac{1}{2}\eta_2[2x\mbox{cos}\theta_1-\mbox{cos}^2\theta_1-2x^2\mbox{cos}\theta_1\mbox{cos}\theta_2/(\mbox{cos}^2\theta_1+\mbox{cos}^2\theta_2)]=f(x)$,
where we have assumed $x=\sqrt{\eta_1/\eta_2}$. Since
$f(\mbox{cos}\theta_1)=\eta_2\mbox{cos}^2\theta_1(\mbox{cos}\theta_1-\mbox{cos}\theta_2)^2/[2(\mbox{cos}^2\theta_1+\mbox{cos}^2\theta_2)]
\ge 0$ and
$f(Ra_1)=\eta_2\mbox{cos}^2\theta_1(\mbox{cos}\theta_1-\mbox{cos}\theta_2)^2/[2(\mbox{cos}\theta_1+\mbox{cos}\theta_2)^2]
\ge 0$, one immediately sees that $P \le P_{Ra}$ for any
$\sqrt{\eta_1/\eta_2}$ in $[\mbox{cos}\theta_1, Ra_1]$. These
observations indicate that the bound presented in this work is
independent of those of former works, and sometimes it can provide
tighter bound of the total success probability $P$, as we have
expected.

\section{Conclusion}
To summarize, we have proposed a general description on the UD of
mixed states from the system-ancilla model, and presented a
procedure to reduce this to a standard SDP problem, which makes
the problem to be solvable numerically. On the UD between two
mixed states, we have introduced the canonical vectors and partly
reduced the original problem to the UD between pairs of canonical
vectors. We present a series of new upper bounds on the total
success probability which depends on both the ratio of the prior
probabilities and the input state structures. This indicates that
the results in \cite{bergou} are universal for any type of input
states. It also should be mentioned that throughout the paper we
mainly concentrate on the diagonal elements of the corresponding
matrices. In practice, the non-diagonal elements also play
important roles which deserves further investigation.

\section{Acknowledgement}
The authors thank M. Kleinmann for very valuable comments and
suggestions and for bringing to our attention earlier work about
simultaneous decomposition of two mixed states. This work was
funded by the National Fundamental Research Program, the National
Natural Science Foundation of China (Grant No. 10674127), the
Innovation Funds from the Chinese Academy of Sciences, and Program
for New Century Excellent Talents in University.

\end{document}